\documentclass[10pt,letterpaper,aps,prb,twocolumn,superscriptaddress,showpacs]{revtex4-1}
\usepackage{graphicx}
\usepackage{epsfig}
\usepackage{dcolumn}
\usepackage{bm}
\usepackage{times}
\usepackage{mathptmx}
\usepackage{color}
\usepackage{amsmath,amssymb}
\usepackage{xfrac}
\usepackage{amsfonts}
\usepackage{xspace}
\begin{document}

\title{Terahertz time-domain spectroscopy of transient metallic and superconducting states}

\author{J.\ Orenstein}
\email{jworenstein@lbl.gov}
\affiliation{Department of Physics, University of California, Berkeley CA 94720, USA}
\affiliation{Materials Science Division, Lawrence Berkeley National Laboratory, Berkeley CA 94720, USA}
\author{J.\ S.\ Dodge}
\email{jsdodge@sfu.ca}
\affiliation{Department of Physics, Simon Fraser University, Burnaby, British Columbia, V5A 1S6, Canada}
\affiliation{Canadian Institute for Advanced Research, Toronto, Ontario, M5G 1Z8, Canada}

\begin{abstract}
Time-resolved terahertz time-domain spectroscopy (THz-TDS) is an ideal tool for probing photoinduced nonequilibrium metallic and superconducting states.  Here, we focus on the interpretation of the two-dimensional response function $\Sigma(\omega;t)$ that it measures, examining whether it provides an accurate snapshot of the instantaneous optical conductivity, $\sigma(\omega;t)$. For the Drude model with a time-dependent carrier density, we show that $\Sigma(\omega;t)$ is not simply related to $\sigma(\omega;t)$. The difference in the two response functions is most pronounced when the momentum relaxation rate of photocarriers is long, as would be the case in a system that becomes superconducting following pulsed photoexcitation. From the analysis of our model, we identify signatures of photoinduced superconductivity that could be seen by time-resolved THz-TDS.
\end{abstract}

\maketitle

\section{Introduction}
\label{sec:intro}

Time-domain terahertz spectroscopy (THz-TDS) probes the optical conductivity of metals and superconductors by measuring the current transient induced by a subpicosecond electric field pulse.  Time-resolved THz-TDS exploits the short duration of the THz probe to detect how the conductivity changes in response to pulsed photoexcitation after a controlled delay---providing a way to take snapshots of the optical conductivity with picosecond time resolution.  This scheme has been used effectively to measure transient photoconductivity in a wide variety of bulk and nanostructured semiconductor systems,\cite{Ulbricht:2011} and its range of application is growing.\cite{Averitt:2002,Basov:2011,Lloyd-Hughes:2012}  Recently, THz-TDS spectra of the photoexcited normal state of high-T$_c$ cuprates and K$_3$C$_{60}$ have been presented as evidence of transient photoinduced superconductivity, because they resemble the equilibrium conductivity spectra obtained well below the superconducting transition temperature.~\cite{Fausti:2011,Hu:2014,Kaiser:2014,Mitrano:2015} However, the straightforward interpretation of THz-TDS as a snapshot of the optical conductivity spectrum breaks down when the characteristic relaxation times are comparable to the photoexcitation delay, because the Fourier transform involves times that precede photoexcitation.\cite{Averitt:2000,Nemec:2002,Hendry:2005,Nemec:2005,Nemec:2005a,Nienhuys:2005,Schins:2007,Schins:2011,Ulbricht:2011}

In order to assess the importance of this issue for interpreting measurements on transient metallic and superconducting states, we analyze a simple model of transient photoconductivity.\cite{Nemec:2005,Nemec:2005a,Nienhuys:2005,Schins:2007}   For photoexcitation at $t=0$, we compare the time-dependent instantaneous conductivity change, $\delta\sigma(\omega;t)$, with the response function $\Sigma(\omega;t)$ measured in time-resolved THz-TDS. We find that $\Sigma$ shows large-amplitude deviations from $\delta\sigma$ when the photocarrier Drude scattering time is longer than $t$,\cite{Nienhuys:2005} a regime that may be accessible in a transient photoinduced superconductor. Our analysis offers improved guidance on how and when time-resolved THz-TDS spectra can be interpreted as a conductivity snapshot of a transient state.

\begin{figure}
\begin{center}
\includegraphics[width=0.8\columnwidth]{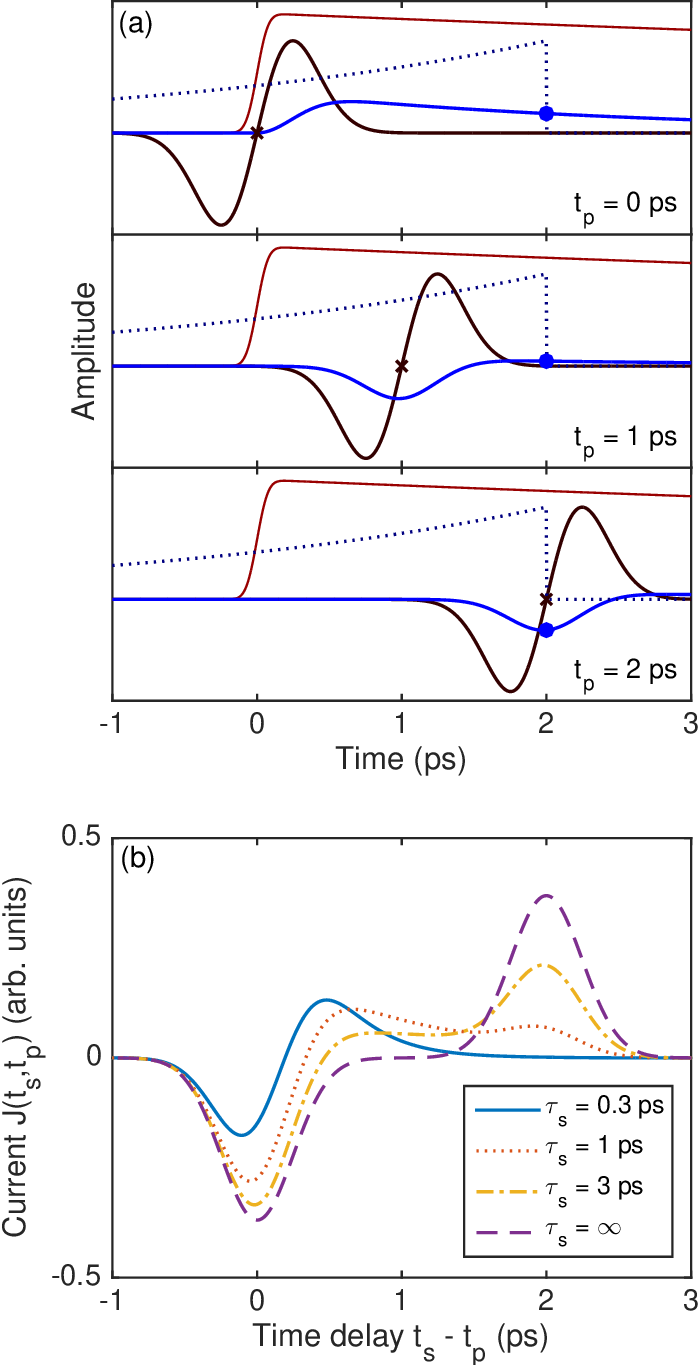}
\end{center}
\caption{(Color online) (a) Schematic illustration of the pulse sequence and time delays in a time-resolved THz-TDS measurement performed on the model system described in the text. The red lines represent the photoexcited carrier density that is generated by a pump pulse at $t=0$, then decays on a timescale $1/\Gamma = 20$~ps. The black lines represent the electric field of the THz probe pulse, shown for increasing time delay, $t_p$ (marked by $\times$), in the three panels. Blue lines depict the corresponding nonequilibrium current, and blue dots mark its value at $t_s = 2$~ps. The dotted line shows the impulse response function $\sigma(t_s - t)$ for a Drude metal. The truncation of the impulse response at $t=0$ differentiates $\Sigma(\tau,t_s)$ from $\delta\sigma(\tau,t_s)$. (b) Nonequilibrum current as a function of $t_s-t_p$ for fixed $t_s=$2 ps and $1/\Gamma = 20$~ps. With increasing momentum relaxation time, $\tau_s$, a pulse of nonequilibrium current centered on $t_p=0$ grows in amplitude.}
\label{fig:jtime}
\end{figure}

We consider a pump-probe experiment in which a material is photoexcited by a strong pump pulse with intensity profile $I(t)\approx I_0\delta(t)$, and the current induced by a THz-frequency probe field is measured at time $t$. The photoinduced change in the current is then
\begin{equation}
\delta J(t) = \iint_{-\infty}^{\infty} E(t-\tau)I(t-\tau_e)\sigma^{(3)}(\tau,\tau_e)\,d\tau\,d\tau_e,
\label{eq:djsigma3}
\end{equation}
where $E(t - \tau)$ is the electric field of the THz probe at the moment $\tau$ before the observation time, and $\sigma^{(3)}(\tau,\tau_e)$ is a third-order susceptibility, in sense that its associated current is proportional to both the THz probe field and the pump intensity (two powers of field). By defining
\begin{equation}
\Sigma(\tau,t) = \int_{-\infty}^{\infty} I(t-\tau_e)\sigma^{(3)}(\tau,\tau_e)d\tau_e,
\label{eq:Sigma}
\end{equation}
we can rewrite Eq.~(\ref{eq:djsigma3}) as
\begin{equation}
\delta J(t) = \int_{-\infty}^{\infty} E(t-\tau)\Sigma(\tau,t)d\tau,
\label{eq:djSigma}
\end{equation}
which has the usual linear response form---except, crucially, that the two-dimensional response function $\Sigma(\tau,t)$ has an implicit dependence on $I(t)$ and lacks time invarance.\cite{Averitt:2000,Nemec:2002,Hendry:2005} Kindt and Schmuttenmaer (KS) pointed out that $\Sigma(\tau,t)$ could be readily obtained with time-resolved THz-TDS by independently controlling the delay of the THz probe field with respect to photoexcitation, as illustrated in Fig.~\ref{fig:jtime} and described more thoroughly in Sec.~\ref{sec:THzTDS}.\cite{Kindt:1999}

With a measurement of $\Sigma(\tau,t)$, it is straighforward to compute the Fourier transform $\Sigma(\omega;t)$. The question that arises then is the following: whether, or in what limit, can $\Sigma(\omega;t)$ be considered equivalent to the pump-induced change in the instantaneous optical conductivity, $\delta\sigma(\omega;t)$?\,\cite{Averitt:2000,Nemec:2002,Hendry:2005,Nemec:2005,Nemec:2005a,Nienhuys:2005,Schins:2007,Schins:2011,Ulbricht:2011}  Indeed, whether a response function $\delta\sigma(\omega;t)$ exists at all for an arbitrary nonequilibrium system is itself a problem, one we do not consider here. Instead, we consider a specific example of a class of optically pumped systems for which the concept of an instantaneous linear response function should be applicable. Following photoexcitation, this class of systems passes through a continuous sequence of quasi-equilibrium states, in which the distribution functions of electrons, phonons, magnons, etc.\ can be described by quasi temperatures and chemical potentials. For such systems it should be possible to define the response function $\delta\sigma(\omega;t)$ as the change in the equilibrium $\sigma(\omega)$ that would be measured if the quasi-equilibrium state at time $t$ were entirely metastable.  Below we show by example that although $\delta\sigma(\omega;t)$ can be well defined, it is not, in general, equivalent to the function $\Sigma(\omega;t)$ measured by time-resolved THz-TDS.

\section{Example of photocarriers with Drude response}
\label{sec:s3example}

We treat the Drude model for nonequilibrium photocarriers, but it is helpful to start with elementary equilibrium relationships. For a system of $n$ carriers with Drude scattering rate $\gamma$, the current is given by
\begin{equation}
J(t) = \int_{-\infty}^{\infty} E(t-\tau)\sigma(\tau)d\tau,
\end{equation}
with impulse response
\begin{equation}
\sigma(\tau)=\frac{ne^2}{m}\Theta(\tau)\exp(-\gamma\tau).
\label{eq:DrudeTime}
\end{equation}
The Drude conductivity spectrum is then just the Fourier transform of Eq.~(\ref{eq:DrudeTime}):
\begin{equation}
\sigma(\omega)=\frac{ne^2}{m}\frac{1}{\gamma-i\omega}.
\label{eq:DrudeFreq}
\end{equation}

We turn now to the nonequilibrium case. If photoexcitation creates $\delta n(0)$ carriers that recombine at a rate $\Gamma$, then the instantaneous conductivity is given by the Drude formula with $n(t) = \delta n(0)\Theta(t)\exp(-\Gamma\,t)$,
\begin{equation}
\delta\sigma(\omega;t)=\Theta(t)\frac{\delta n(0)e^2}{m}\exp(-\Gamma\, t)\frac{1}{\gamma-i\omega},
\end{equation}
or, in the time domain,
\begin{equation}
\delta\sigma(\tau,t)=\Theta(t)\frac{\delta n(0)e^2}{m}\exp(-\Gamma\, t)\Theta(\tau)\exp(-\gamma\tau).
\label{eq:dsigmataut}
\end{equation}
We emphasize here that in both $\delta\sigma(\tau,t)$ and $\delta\sigma(\omega;t)$, the dependence on $t$ is entirely through the state variable $n(t)$, which is then assumed constant when we consider the dependence of $\delta\sigma$ on the dynamical variables $\tau$ and $\omega$.

By contrast, $\Sigma(\tau,t)$ has an explicit dynamical dependence on $t$. Because of the relative simplicity of our model, we can determine $\Sigma(\tau,t)$ directly by integrating the classical equation of motion for the photocarriers. As shown in Appendix~\ref{sec:appendix}, we obtain\cite{Nemec:2005}
\begin{equation}
\Sigma(\tau,t)=\Theta(t-\tau)\Theta(t)\frac{\delta n(0)e^2}{m}\exp(-\Gamma\,t)\Theta(\tau)\exp(-\gamma\tau).
\label{eq:SigmaEOM}
\end{equation}
Eq.~(\ref{eq:SigmaEOM}) is identical to Eq.~(\ref{eq:dsigmataut}) except for the additional Heaviside function, $\Theta(t-\tau)$, which prevents carriers from contributing to the integrand in Eq.~(\ref{eq:djSigma}) before they are created at $t=0$ (or $\tau=t$).

Substituting this form for $\Sigma(\tau,t)$ into Eq.~(\ref{eq:djSigma}) yields,
\begin{multline}
\delta J(t)=\Theta(t)\frac{\delta n(0)e^2}{m}\exp(-\Gamma\, t)\\
\times\int_{-\infty}^{\infty} E(t-\tau)\Theta(t-\tau)\Theta(\tau)\exp(-\gamma\tau).
\label{eq:djSigmaTime}
\end{multline}
In the following section we show how time-resolved THz-TDS is applied to measure $\Sigma(\tau,t)$, and compare $\Sigma(\omega;t)$ with $\delta\sigma(\omega;t)$.

\section{Time-resolved terahertz spectroscopy}
\label{sec:THzTDS}

As a preliminary step we review the methodology of THz-TDS as applied to the equilibrium optical conductivity.\cite{Ulbricht:2011} THz-TDS effectively measures the current at a time $t_s$ that is induced by a THz electric field pulse, $E(t)=E_p(t-t_p)$, centered on a time $t_p$:
\begin{equation}
\hat{J}(t_s,t_p)=\int_{-\infty}^\infty E_p(t_s-t_p-\tau)\sigma(\tau)d\tau.
\label{eq:jTHzTDS}
\end{equation}
The induced current $\hat{J}(t_s,t_p)$ is inferred from a time-resolved measurement of the electric field reflected from, or transmitted through, a medium under test. It depends only on the difference $t_s - t_p$ between the sampling and probe arrival times, so it can be measured by scanning either $t_s$ or $t_p$. Using the convolution theorem, $\sigma(\omega)$ can be obtained by Fourier transforming $\hat{J}(t_s,t_p)$ along the $\Delta\,t = t_s - t_p$ direction.

Time-resolved THz-TDS focuses on the change in the response to the THz probe that is induced by a pump pulse.  The sequencing of pump and THz probe pulses is illustrated in Fig.~\ref{fig:jtime}(a). Using Eq.~(\ref{eq:djSigmaTime}), the nonequilibrium current for Drude photocarriers is
\begin{multline}
\delta \hat{J}(t_s,t_p)=\Theta(t_s)\frac{\delta n(0)e^2}{m}\exp(-\Gamma\,t_s)\\
\times\int_{-\infty}^{\infty} E_p(t_s-t_p-\tau)\Theta(t_s-\tau)\Theta(\tau)\exp(-\gamma\,\tau)d\tau.
\label{eq:djtstp}
\end{multline}
Note that $\delta \hat{J}(t_s,t_p)$ is now a function of both the sampling and probe arrival times, rather than just their difference, as in equilibrium THz-TDS; this reflects the breaking of time-invariance by the pump. An example of the transient nonequilibrium current calculated from Eq.~(\ref{eq:djtstp}) is illustrated in Fig.~\ref{fig:jtime}(b), which shows $\delta \hat{J}(t_s,t_p)$ as a function of $t_s-t_p$ with $t_s=$ 2 ps, for several values of $\tau_s\equiv 1/\gamma$. For $\tau_s = 0.3$~ps, the current follows the electric field pulse, with a lag and slight distortion caused by convolution with the Drude impulse response. As $\tau_s$ increases, a feature emerges near $t_p = 0$ that has a completely different origin, illustrated in the $t_p = 0$ panel of Fig.~\ref{fig:jtime}(a): here, only part of the THz field can induce current, creating an unbalanced current impulse that persists until the measurement time $t_s$. When $\tau_s\rightarrow\infty$, as in a superconductor, these two features have equal and opposite magnitude; however, their separation in time will vary with the observation time $t_s$.

The procedure introduced by KS is to scan $t_p$ at fixed $t_s$, such that the integral expression for the current retains the form of Eq.~(\ref{eq:jTHzTDS}), but with the two-dimensional nonequilibrium response function $\Sigma(\tau,t_s)$ in place of $\sigma(\tau)$, as in Eq.~(\ref{eq:djSigma}). If we now Fourier transform $\Sigma(\tau,t_s)$ with respect to $\tau$, we get
\begin{multline}
\Sigma(\omega;t_s)=\Theta(t_s)\frac{\delta n(0)e^2}{m}\exp(-\Gamma\,t_s)\frac{1}{\gamma-i\omega}\\
\times\left[1-\exp(-\gamma\,t_s)\exp(i\omega t_s)\right],
\label{eq:SigmaTHzTDS}
\end{multline}
which differs from the instantaneous conductivity $\delta\sigma(\omega;t_s)$ in Eq.~(\ref{eq:dsigmataut}) by the term in square brackets. This result is consistent with earlier results on the nonequilibrium Drude model,\cite{Hendry:2005,Nemec:2005,Nemec:2005a,Nienhuys:2005,Schins:2007} expressed in a way that allows more immediate comparison with experiment.

We have assumed impulsive excitation, $I(t)\approx I_0\delta(t)$, that causes the conductivity to change abruptly. When the excitation pulse width $\tau_w$ cannot be neglected, we expect the oscillations in $\Sigma(\omega;t_s)$ to become damped as $\omega\tau_w \gtrsim 1$. Another factor that can lead to damping occurs if the photoinduced change in conductivity has a finite risetime.  This would be reflected in the dependence of the third-order susceptibility $\sigma^{(3)}(\tau,\tau_e)$ on $\tau_e$ in Eq.~(\ref{eq:Sigma}). In the presence of either form of broadening of the step-function change in conductivity, the deviation of $\Sigma$ from $\sigma$ will be most pronounced at low frequencies, which is indeed observed.\cite{Averitt:2000,Beard:2000}

\section{Discussion}

Eq.~(\ref{eq:SigmaTHzTDS}) greatly clarifies the conditions under which the time-resolved THz-TDS spectrum $\Sigma(\omega;t_s)$ approximates the instantaneous linear response, $\delta\sigma(\omega;t_s)$. First, it shows that photocarrier recombination simply rescales the overall spectrum by $\exp(-\Gamma t_s)$, so the measurement fidelity is not fundamentally limited by the recombination time. The critical parameter is $\gamma\, t_s$, the product of the momentum relaxation rate and the sampling time.

\begin{figure}
\begin{center}
\includegraphics[width=0.8\columnwidth]{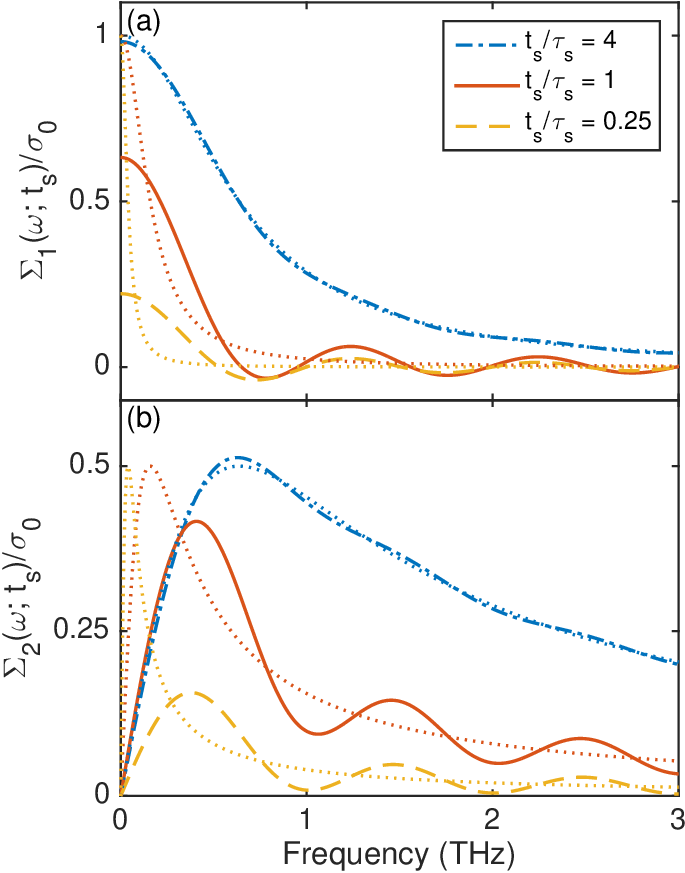}
\end{center}
\caption{(Color online) (a) Real and (b) imaginary part of $\Sigma(\omega;t_s)$ as a function of $\omega$ for several values of $\tau_s$, with $t_s$ fixed at 1~ps and $\Gamma = 0$. The Drude conductivity for same values of $\tau_s$ is shown as dotted lines for comparison.  Spectra with different $\tau_s$ are normalized to $\sigma_0 = n(0)\tau_s e^2/m$ to illustrate the variation in frequency dependence. When $t_s/\tau_s$ is large $\Sigma(\omega;t_s)$ asymptotically approaches the instantaneous (Drude) conductivity. For $\tau_s\geq t_s$ the temporal cutoff at $t=0$ generates oscillations, with period 2$\pi/t_s$, along the frequency axis of $\Sigma(\omega;t_s)$.}
\label{fig:SigmaTauS}
\end{figure}

The crossover in the nature of the spectra at $\gamma\, t_s\sim 1$ is illustrated Fig.~\ref{fig:SigmaTauS}(a), in which we plot $\Sigma(\omega;t_s)$ as a function of $\omega$ for several values of $\tau_s$, with $t_s$ fixed at 2~ps. The instantaneous Drude conductivity $\delta\sigma(\omega;t_s)$ for the same values of $\tau_s$ is shown as dotted lines for comparison, and spectra with different $\tau_s$ are normalized to $\delta\sigma(\omega;t_s=0)$ to illustrate the variation in frequency dependence.

When $t_s/\tau_s$ is large, the component of the current sampled at $t_s$ from carriers accelerated at $t=0$ is exponentially small. Consequently, $\Sigma(\omega;t_s)$ asymptotically approaches the instantaneous conductivity, which has the Drude form in our simple example. On the other hand, if $\tau_s$ is comparable to or greater than $t_s$, a component of the current that would be present if the state were metastable is cutoff at $t=0$.  The temporal cutoff generates oscillations with period 2$\pi/t_s$ along the frequency axis of $\Sigma(\omega;t_s)$, which is clearly no longer simply related to $\delta\sigma(\omega;t_s)$.

\begin{figure}
\begin{center}
\includegraphics[width=\columnwidth]{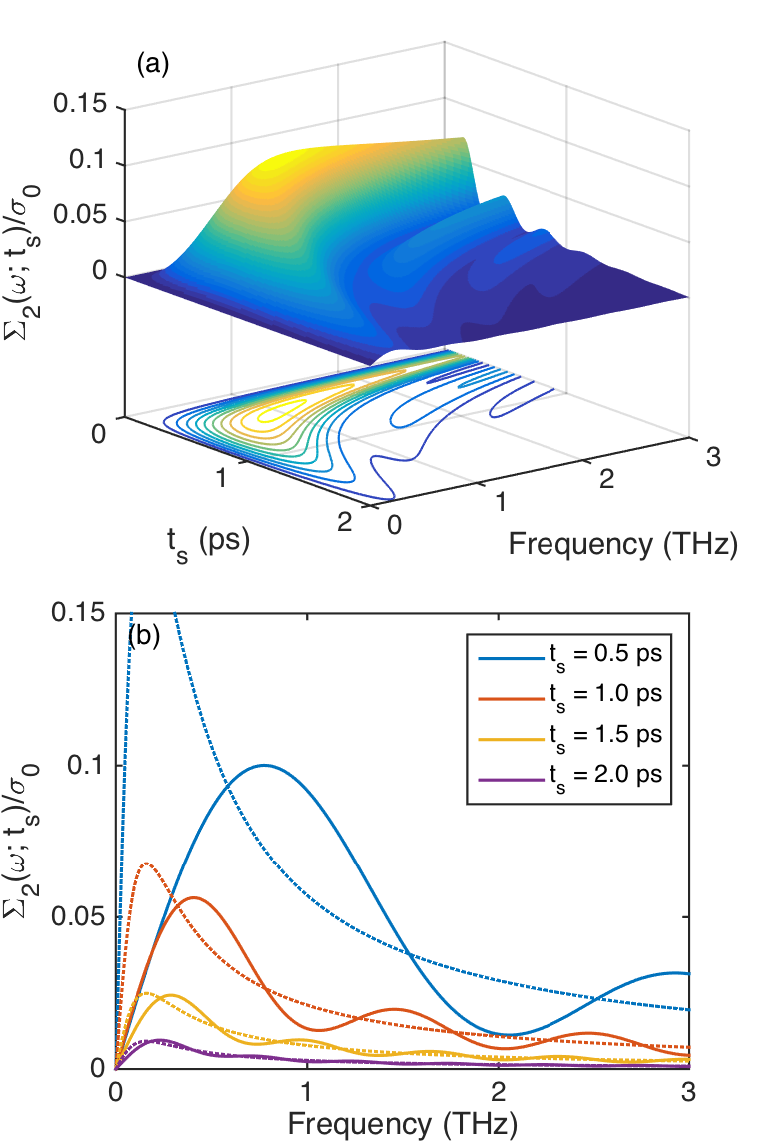}
\end{center}
\caption{(Color online) Simulation of time-resolved THz-TDS response function for photocarriers whose dynamics are characterized by a time-independent scattering rate. (a) Imaginary part of $\Sigma(\omega;t_s)$, with $\tau_s =1$~ps and $1/\Gamma=0.5$~ps, in both 3D and contour plot representations. Contour levels indicate 0.01 steps from 0.01 to 0.1, inclusive. (b) Spectra of $\Sigma_2(\omega;t_s)$ (solid lines) for several values of $t_s$, with $\delta\sigma_2(\omega; t_s)$ (dotted lines) shown for comparison. The spectrum for each $t_s$ is denoted by color in the legend; both $\Sigma_2$ and $\sigma_2$ decrease with increasing $t_s$, as the carriers decay. All spectra are normalized to $\sigma_0 \equiv \delta n(0)e^2\tau_s/m$.}
\label{fig:SigmaTs}
\end{figure}

Fig.~\ref{fig:SigmaTs} shows this behavior in more detail as it would appear for photocarriers with $\tau_s = 1$~ps and $1/\Gamma = 0.5$~ps. The imaginary part of $\Sigma(\omega;t_s)$ has oscillations that appear as ridges along constant values of $\omega t_s$, with an amplitude that decays exponentially with $\gamma\, t_s$.  Similar oscillations were observed previously in both measurements\cite{Beard:2000} and simulations,\cite{Nienhuys:2005} and Eq.~\ref{eq:SigmaTHzTDS} clarifies their origin.

\begin{figure}
\begin{center}
\includegraphics[width=\columnwidth]{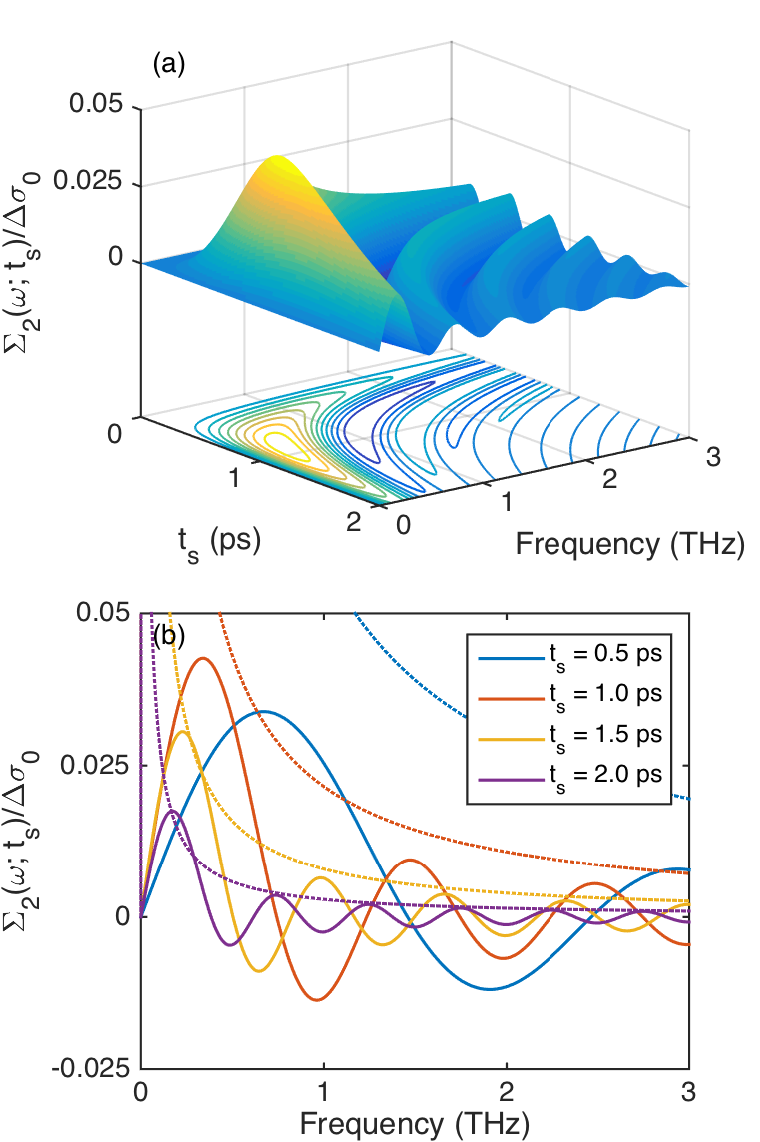}
\end{center}
\caption{(Color online) Simulation of time-resolved THz-TDS response function for a model of transient photoinduced superconductivity. (a) Imaginary part of $\Sigma(\omega;t_s)$, with $\tau_s = 1$~ps and $1/\Gamma=0.5$~ps, in both 3D and contour plot representations. Contour levels indicate 0.005 steps from -0.01 to 0.04, inclusive. (b) Spectra of $\Sigma_2(\omega;t_s)$ (solid lines) for several values of $t_s$, with $\delta\sigma_2(\omega; t_s)$ (dotted lines) shown for comparison. The spectrum for each $t_s$ is denoted by color in the legend; with increasing $t_s$, $\sigma_2$ decreases uniformly as the carriers decay, while $\Sigma_2$ shows oscillations about $\Sigma_2 \approx 0$, with period $2\pi/t_s$ and an amplitude that decays with both $\omega$ and $t_s$. All spectra are normalized to $\Delta\sigma_0 \equiv \Delta ne^2\tau_s/m$.}
\label{fig:SigmaTsTF}
\end{figure}

Finally, we consider what might be observed in a time-resolved THz-TDS measurement in which a transient superconducting phase is generated at $t=0$ by a laser pulse. Developing a phenomenological description of transient superconductivity is clearly not as straightforward as modeling a transient photoconductor.  One approach that is directly amenable to our analysis is based on the two-fluid model, which describes the current response in terms of normal fluid and superfluid components, with spectral weights $n_ne^2/m$ and $n_se^2/m$ respectively.  The normal fluid conductivity is described by the Drude spectrum, while the superfluid component is characterized by an infinite momentum relaxation time.  We can then formulate photoinduced superconductivity as the generation of superfluid spectral weight $\Delta ne^2/m$ by transfer from the normal fluid.  If we assume that the photoinduced superfluid has a lifetime $1/\Gamma$, then
\begin{multline}
\Sigma(\omega;t_s)=i\Theta(t_s)\frac{\Delta n\,e^2}{m}\exp(-\Gamma\,t_s)\\
\times\left\{\frac{1}{\omega}[1-\exp(i\omega\, t_s)]-\frac{1}{\tilde{\omega}}[1-\exp(i\tilde{\omega}\, t_s)]\right\},
\label{eq:SigmaTF}
\end{multline}
with $\tilde{\omega}\equiv \omega + i\gamma$. Fig.~\ref{fig:SigmaTsTF} illustrates the spectra predicted by Eq.~(\ref{eq:SigmaTF}) for several values of $t_s$, with $\tau_s=$1~ps and $\tau_r=$0.5~ps. The spectral shape is dominated by underdamped oscillations that originate from the sharp cutoff in the time-domain response at $t=0$, as discussed above. We note that in this description of transient superfluidity, $\Sigma(\omega;t_s)$ never approaches $\delta\sigma(\omega;t_s)$ because of the undamped contribution to $\hat{J}(t_s,t_p)$ from super carriers generated at $t=0$. With an overall sign change, Eq.~(\ref{eq:SigmaTF}) should also describe $\Sigma(\omega;t_s)$ when photoexcitation suppresses superconductivity, for example by transferring spectral weight from the superconducting carriers to quasiparticles above the gap.~\cite{Averitt:2000}

As mentioned above, different phenomenological descriptions of a transient superconducting state are possible.  For example, rather than generating a fully coherent superfluid component, one could imagine that the effect of the pump pulse is to suddenly increase the momentum relaxation time of the entire electron fluid to some large but still finite value. The system would subsequently return to equilibrium through the decay of partial coherence and recovery of the normal state $\tau_s$.  We describe the TD-THz response for the case of a time-varying momentum relaxation rate in the Appendix, where we obtain the response function $\Sigma(\tau,t)$ in terms of $\gamma(t)$. Unlike the models considered above, we believe that the instantaneous conductivity, $\sigma(\omega;t)$, is not well-defined in a system where $\gamma$ depends explicitly on the time. Still, $\Sigma(\omega;t)$ remains a valid response function, and can exhibit features that are similar to those predicted for the two-fluid model. In this case, the deviations from a Drude spectrum will smaller than in the fully coherent superconductor, particularly if the maximum $\tau_s$ reached by the partially coherent state does not exceed the sampling time.

\section{Summary and conclusions}

Time-domain terahertz spectroscopy provides an elegant method for doing time-resolved photoconductivity measurements.  In this paper we focused on whether the response function $\Sigma(\omega;t_s)$ that is typically measured in time-resolved THz-TDS can be interpreted as the photoinduced change $\delta\sigma(\omega;t_s)$ in the instantaneous optical conductivity. Within a simple model, we showed that $\Sigma(\omega;t_s)$, is never equivalent to $\delta\sigma(\omega;t_s)$, although we also found that $\Sigma(\omega;t_s)$ approaches $\delta\sigma(\omega;t_s)$ asymptotically in the limit $\gamma\, t_s\rightarrow \infty$, where $\gamma$ is the Drude relaxation rate of the nonequilibrium carriers. In this limit, the current measured at $t_s$ has an exponentially small dependence on the field applied at $t<0$; that is, before the pump pulse arrives. In the opposite regime, $\gamma\, t_s\lesssim 1$, we presented an analytic expression that shows that $\Sigma(\omega;t_s)$ and $\delta\sigma(\omega;t_s)$ are entirely distinct response functions for the nonequilibrium Drude model. Here, the absence of nonequilibrium carriers for $t<0$ creates a current imbalance for fields applied near $t=0$ that persists until the measurement time. Nevertheless, we believe that even in this regime, information about the number density and mobility of the photocarriers can be obtained by comparing $\Sigma(\omega;t_s)$ with theoretical models.\cite{Nemec:2005,Nemec:2005a,Nienhuys:2005,Schins:2007,Schins:2011}

The effort to better understand the relationship of $\Sigma(\omega;t_s)$ to $\delta\sigma(\omega;t_s)$ was largely motivated by experiments reporting photoinduced transient superconductivity in cuprates and in K$_3$C$_{60}$. The evidence presented for superconductivity is that the instantaneous conductivity following photoexcitation in the normal state, $\sigma(\omega;t)$, has features characteristic of the equilibrium $\sigma(\omega)$ measured at $T\ll T_c$. To obtain the instantaneous conductivity it is assumed that $\sigma(\omega;t)=\sigma(\omega)+\Sigma(\omega;t)$, which is based on regarding $\Sigma(\omega;t)$ and $\delta \sigma(\omega;t)$ as equivalent. However, we have shown that these response functions are not equivalent, and differ most strongly when carrier momentum relaxation rates become long. Superconductivity, in which the condensate momentum relaxation time diverges, is the most extreme example of the inequivalence of $\Sigma(\omega;t)$ and $\delta\sigma(\omega;t)$.

We considered two perspectives in attempting to model the response function $\Sigma(\omega;t)$ appropriate to photoinduced transient superconductivity. For a two-fluid model with a transient, fully coherent superfluid component, we predict strong oscillations along the frequency axis of $\Sigma(\omega;t)$, with a period inversely related to the sampling time. In the second perspective, the pump induces a partially coherent state, with an enhanced, but still finite, momentum relaxation time. In this model, deviations from a Drude spectrum are again expected, though damped by the limited coherence time. In either case, the instantaneous conductivity $\sigma(\omega;t)$ is unobservable or ill-defined in the most physically interesting regimes, while  time-resolved THz-TDS measures $\Sigma(\omega;t)$ directly. To advance research on photoinduced superconductivity and other collective states, we believe it important to distinguish them.

\begin{acknowledgments}
JSD thanks J. Bechhoefer for suggesting the equation of motion approach presented in the appendix, and acknowledges support from NSERC and CIFAR. JO acknowledges the Office of Basic Energy Sciences, Materials Sciences and Engineering Division, of the U.~S.\ Department of Energy under Contract No.~DE-AC02-05CH11231 for support.
\end{acknowledgments}

\appendix
\section{Classical derivation of the nonequilibrium response}
\label{sec:appendix}
To extend the classical Drude model to photoexcited materials, we let both the carrier density $n$ and the damping rate $\gamma$ depend on time in the usual equation of motion. The current is then related to the field through the linear, first-order differential equation
\begin{equation}
\frac{dJ}{dt} + \left(\gamma - \frac{1}{n}\frac{dn}{dt}\right)J = \frac{ne^2}{m}E.
\label{eq:eom}
\end{equation}
The extra damping term follows from the chain rule with $J = nev$, and causes the current to decay more rapidly when the carrier density decreases, as expected. Conversely, the damping term decreases when the carrier density increases, because our model incorrectly assumes that all carriers move with the same velocity.\cite{Nienhuys:2005} Others have addressed this problem by expressing the current in terms of a distribution function, but their results reproduce Eq.~(\ref{eq:eom}) in the usual case of carrier decay.\cite{Nemec:2005,Nienhuys:2005}

To solve Eq.~(\ref{eq:eom}), we introduce the integrating factor
\begin{equation}
F(t,t_i) \equiv \frac{n(t_i)}{n(t)}\exp\left[\bar{\gamma}(t_i,t)\,(t-t_i)\right],
\label{eq:intfac}
\end{equation}
with
\begin{equation}
\bar{\gamma}(t_1,t_2) \equiv \frac{\int_{t_1}^{t_2} dt'\gamma(t')}{t_2-t_1}
\label{eq:avdamp}
\end{equation}
the average damping rate over the interval $(t_1,t_2)$. Assuming $n\geq 0$ over $(t_i,t)$, we multiply Eq.~(\ref{eq:eom}) by (\ref{eq:intfac}) and integrate to get
\begin{equation}
J(t) - J(t_i) = \frac{n(t)e^2}{m}\int_{t_i}^t dt' \exp\left[-\bar{\gamma}(t',t)\,(t-t')\right]E(t').
\label{eq:eomsol}
\end{equation}
When $n$ and $\gamma$ are constant, Eq.~(\ref{eq:eomsol}) gives the conventional Drude response. When $n$ and $\gamma$ vary with time, the current $J(t)$ includes contributions from impulses at earlier times $t'$, exponentially weighted by the average damping rate experienced over its history. An impulse with carrier density $n(t')$ will decay by a factor $n(t)/n(t')$ before it contributes to the current $J(t)$, so only the overall factor $n(t)$ appears outside the integral.

Referring now to Eq.~(\ref{eq:djSigma}), we let $n=n_\text{eq}$ and $\gamma = \gamma_\text{eq}$ at equilibrium, and use Eq.~(\ref{eq:eomsol}) to find the current change $\delta J(t)$ following photoexcitation at $t=0$:
\begin{multline}
\Sigma(\tau,t) = \Theta(t)\Theta(t-\tau)\Theta(\tau)\frac{e^2}{m}\left\{n(t)\exp\left[-\bar{\gamma}(t-\tau,t)\tau\right] \right.\\
- \left.n_\text{eq}\exp(-\gamma_\text{eq}\tau)\right\}.
\end{multline}
For the specific case of a photoinduced carrier density $n(t) = n_\text{eq} + \Theta(t)\delta n(0)e^{-\Gamma\,t}$ with a constant Drude scattering rate $\gamma$,
\begin{equation}
\Sigma(\tau,t) = \Theta(t-\tau)\Theta(t)\frac{\delta n(0)e^2}{m}\exp(-\Gamma\,t)\Theta(\tau)\exp(-\gamma\tau).
\label{eq:SigmaTauTapp}
\end{equation}
The second factor of $\Theta(t)$ in Eq.~(\ref{eq:SigmaTauTapp}) is missing in Eq.~(31) of Ref.~\onlinecite{Nemec:2005}, but this appears to be a typographical error, since the factor is necessary to obtain their Eq.~(32). Otherwise, the expressions are equivalent.

\end{document}